\documentclass[aip, pre, twocolumn, superscriptaddress, reprint]{revtex4-1}
\usepackage[T1]{fontenc}
\usepackage{times}
\usepackage{graphicx}

\usepackage{amsfonts, amsmath,amssymb, bm, graphicx}
\usepackage{siunitx}

\usepackage[english]{babel}
\usepackage{lipsum}

\newcommand{\figref}[2]{\hyperref[#1]{\ref{#1}(#2)}}
\usepackage{physics}
\usepackage{lipsum}
\usepackage{changes}

\usepackage[colorlinks=true,allcolors=blue]{hyperref}
\usepackage{footmisc}

\usepackage{upgreek}

\usepackage{soul}
\newenvironment{rcases}
  {\left.\begin{aligned}}
  {\end{aligned}\right\rbrace}

\begin{document}

{
\makeatletter
\def\frontmatter@thefootnote{%
 \altaffilletter@sw{\@fnsymbol}{\@fnsymbol}{\csname c@\@mpfn\endcsname}%
}%

\makeatother
\title{Finite-element dynamic-matrix approach for spin-wave dispersions \\ in magnonic waveguides with arbitrary cross section}

\author{L. K\"orber}\email{l.koerber@hzdr.de}
\affiliation{Helmholtz-Zentrum Dresden - Rossendorf, Institut f\"ur Ionenstrahlphysik und Materialforschung, D-01328 Dresden, Germany}
\affiliation{Fakultät Physik, Technische Universit\"at Dresden, D-01062 Dresden, Germany}

\author{G. Quasebarth}
\affiliation{Helmholtz-Zentrum Dresden - Rossendorf, Institut f\"ur Ionenstrahlphysik und Materialforschung, D-01328 Dresden, Germany}
\affiliation{Fakultät Physik, Technische Universit\"at Dresden, D-01062 Dresden, Germany}

\author{A. Otto}
\affiliation{Fakultät Physik, Technische Universit\"at Dresden, D-01062 Dresden, Germany}


\author{A. Kákay}
\affiliation{Helmholtz-Zentrum Dresden - Rossendorf, Institut f\"ur Ionenstrahlphysik und Materialforschung, D-01328 Dresden, Germany}

\date{\today}

\begin{abstract}
We present a numerical approach to efficiently calculate  spin-wave dispersions and spatial mode profiles in magnetic waveguides of arbitrarily shaped cross section with any non-collinear equilibrium magnetization which is translationally invariant along the waveguide.
Our method is based on the propagating-wave dynamic-matrix approach by Henry \textit{et al.} 
and extends it to arbitrary cross sections using a finite-element method. We solve the linearized equation of motion of the magnetization only in a single waveguide cross section which drastically reduces computational effort compared to common three-dimensional micromagnetic simulations. In order to numerically obtain the dipolar potential of individual spin-wave modes, we present a plane-wave version of the hybrid finite-element/boundary-element method by Frekdin and Koehler~
which, for the first time, we extend to a modified version of the Poisson equation. Our method is applied to several important examples of magnonic waveguides including systems with surface curvature, such as magnetic nanotubes, where the curvature leads to an asymmetric spin-wave dispersion. In all cases, the validity of our approach is confirmed by other methods. Our method is of particular interest for the study of curvature-induced or magnetochiral effects on spin-wave transport but also serves as an efficient tool to investigate standard magnonic problems.
\end{abstract}

\maketitle

\section{Introduction}

Over the last decades, a number of powerful analytical and numerical tools have been developed to describe the linear propagation characteristics of spin waves, the fundamental small-amplitude excitations in magnetically ordered substances. For example, micromagnetic approaches, which treat the magnetization of a solid as a continuous vector field, have been used to derive approximate analytic expressions for the dispersion relation and spatial mode profiles of spin waves with wavelengths in the nano- and micrometer range. These models allow to describe the spin-wave propagation in simple geometries, such as bulk systems, infinitely extended mono- or bilayers \cite{kalinikosTheoryDipoleexchangeSpin1986,ariasExtrinsicContributionsFerromagnetic, gallardoReconfigurableSpinWaveNonreciprocity2019} or waveguides with rectangular cross section \cite{guslienkoEffectiveDipolarBoundary2002a,kostylevDipoleexchangePropagatingSpinwave2007}. Apart from flat geometries, in the emerging field of curvilinear magnetism, which includes studying curvature-induced effects on spin-wave propagation, the dispersion in cylindrical waveguides, magnetic nanotubes with thin mantle \cite{otaloraCurvatureInducedAsymmetricSpinWave2016, otaloraAsymmetricSpinwaveDispersion2017}, curved nanowires \cite{gaidideiLocalizationMagnonModes2018} and narrow ribbons \cite{gaidideiMagnetizationNarrowRibbons2017}, have been described.

Although the developed analytic approaches are versatile and applicable to many different cases, they tend to give only  approximate dispersions and mode profiles. Moreover they are often not available for more complex systems. Therefore, numerical approaches are needed to complement and extend analytical or experimental studies. Dynamic micromagnetic simulations, which rely on a rigorous time integration of the equation of motion of the magnetization on a discrete mesh, have been established as one of the standard tools to study spin-wave propagation numerically \cite{donahueOOMMFUserGuide1999,kakaySpeedupFEMMicromagnetic2010, vansteenkisteDesignVerificationMuMax32014}. In such dynamic simulations, spin waves are typically obtained by exciting the magnetic system with a microwave field pulse, letting the system evolve according to the appropriate equation of motion, later on extracting the spatial mode profiles and computing the wavelengths and frequencies by means of Fourier analysis. This, however, requires an approximate \textit{a priori} knowledge of the mode profiles and does not always allow to separate degenerate modes.

Alternatively, a numeric micromagnetic method known as dynamic-matrix approach can be used to obtain spin-wave mode profiles and frequencies directly by numerically solving a linearized version of the equation of motion of the magnetization  using an appropriate eigensolver \cite{grimsditchMagneticNormalModes2004,daquinoComputationMagnetizationNormal2012}. One advantage of such a method is that it directly yields frequencies and mode profiles without the need for additional post processing. Moreover, degenerate modes as well as modes with non-trivial spatial profile can be resolved. The dynamic-matrix approach has already been implemented successfully to study standing waves in confined magnetic elements, using both finite-element and finite-difference methods of discretization  \cite{giovanniniSpinExcitationsNanometric2004,naletovIdentificationSelectionRules2011,taurelCompleteMappingSpinwave2016,brucknerLargeScaleFiniteElement2019}. Recently, in an excellent work,  Henry \textit{et al.} \cite{henryPropagatingSpinwaveNormal2016} succeeded to extend this dynamic-matrix approach for propagating spin waves in systems with a translational invariant magnetic equilibrium, using plane-wave demagnetization tensors. They employed a finite-difference method to efficiently obtain the spin-wave dispersion by numerically solving the linearized equation of motion only in the cross-section of the magnetic medium perpendicular to the propagation direction, modelling infinite magnetic slabs or thin flat waveguides using rectangular elements. This approach was already successfully applied \textit{e.g.}\ to study spin waves in multilayer systems \cite{grassiSlowWaveBasedNanomagnonicDiode2020} and those propagating along Bloch domain walls \cite{henryUnidirectionalSpinwaveChanneling2019}. Next to the aforementioned benefits, this propagating-wave dynamic-matrix approach allows for almost arbitrary wave-number precision in the propagation direction, something which can become extremely costly when modelling a full three-dimensional magnetic specimen.

Although the finite-difference-based propagating-wave dynamic-matrix approach is very efficient and allows to study \textit{e.g.} complex multilayer systems, it is not suitable for waveguides with arbitrary (bounded) cross section, for example tubular systems, or, in general, systems with surface curvature. The aim of this paper is to provide a complimentary approach to the one presented in Ref.~\citenum{henryPropagatingSpinwaveNormal2016} which allows to study exactly such systems. For this purpose, we employ a different discretization type, namely the finite-element method (FEM), by modeling the cross section of an arbitrary waveguide using triangular elements. As a key difference, instead of using plane-wave demagnetization tensors, we obtain the dipolar potential of the individual spin-wave modes using the Fredkin-Koehler method~\cite{fredkinHybridMethodComputing1990}, which is a hybrid finite-element/boundary-element method to solve the Poisson equation on a finite volume. Here, for the first time, we present an extension of this method for the \textit{screened} Poisson equation, which governs the lateral dipolar field of propagating waves.

To this end, in Sec.~\ref{sec:theoretical-model}, we review the theoretical basis of the eigenvalue problem to calculate spin-wave dispersions within the context of micromagnetism. Following, in Sec.~\ref{sec:components}, the different magnetic interactions considered in the present work are introduced for the case of propagating waves. The numerical implementation, including the plane-wave Fredkin-Koehler method, is described in Sec.~\ref{sec:num-implentation}. Finally, we apply our approach to different waveguide cross sections in Sec.~\ref{sec:applications}. We believe that this work is of particular interest for the emerging field of curvilinear magnonics, but also allows to calculate spin-wave dispersions in various standard magnonic problems.

\section{Theoretical model}\label{sec:theoretical-model}

\subsection{Basics of micromagnetism}

Within the theory of micromagnetism \cite{brownjr.Micromagnetics1963, gurevichMagnetizationOscillationsWaves1996}, the magnetization $\bm{M}(\bm{r},t)$ of a magnetic body under isothermal conditions (far below the Curie temperature of the material at hand) is treated as a continuous vector field, subject to the constraint $\abs{\bm{M}}=M_\mathrm{s}$, with $M_\mathrm{s}$ being the saturation magnetization of the material. At a constant temperature and exposed to some external field $\bm{H}_\mathrm{ext}(\bm{r},t)$, a stable magnetic equilibrium within a given specimen is reached once its Gibb's free energy is at a local minimum. In terms of the normalized magnetization field $\bm{m}=\bm{M}/M_\mathrm{s}$, the Gibb's free energy $\mathcal{G}(\bm{m})$ of a magnetic specimen is given by
\begin{equation}\label{eq:energy}
   \begin{split}
        \mathcal{G}(\bm{m})   = &- \mu_0 M_\mathrm{s}^2\int\limits_V \mathrm{d}V \, \bm{m}(\bm{r},t) \cdot  \bm{h}_\mathrm{ext}(\bm{r},t) \ \\ & -\frac{\mu_0 M_\mathrm{s}^2}{2} \int\limits_V \mathrm{d}V \, \bm{m}(\bm{r},t) \cdot  \bm{h}_\mathrm{int}(\bm{m};\bm{r},t)
   \end{split}
\end{equation}
with $\mu_0$ being the vacuum permeability, $V$ the volume of the magnetic body,  $\bm{h}_\mathrm{ext}=\bm{H}_\mathrm{ext}/M_\mathrm{s}$ the unitless external field
and $\bm{h}_\mathrm{int}$ the unitless internal magnetic field produced solely by the magnetization itself \footnote{The term \textit{internal} here is meant in a thermodynamic sense (as in \textit{internal energy}) and not in a spatial sense, because obviously the external field is also present within the volume of the sample. In a spatial sense, the internal field would be the effective field defined in Eq.~\eqref{eq:full-eff-field}.}. We have chosen the unitless representation of the magnetization and field in anticipation of the numerical implementation. Moreover, we assume to have a homogeneous saturation $M_\mathrm{s}$ in the material. Extending this method to inhomogeneous materials does not require any difficult steps.

For most common magnetic interactions, the internal field can be expressed as being linear in the magnetization,
\begin{equation}\label{eq:lin-eff-field}
    \bm{h}_\mathrm{int}(\bm{m};\bm{r},t)  =  - \hat{\mathbf{N}}\bm{m}(\bm{r},t)
\end{equation}
using the self-adjoint operator $\hat{\mathbf{N}}$ which describes the magnetic self interactions \footnote{The negative sign in the internal field is usually chosen by convention in accordance with the demagnetizing/dipolar field}. This includes the symmetric exchange interaction, the dipolar interaction, and uniaxial crystal anisotropy,
\begin{equation}\label{eq:interactions}
    \hat{\mathbf{N}} = \hat{\mathbf{N}}^{\mathrm{(exc)}} +\hat{\mathbf{N}}^{\mathrm{(dip)}} +  \hat{\mathbf{N}}^{\mathrm{(uni)}} + \hdots
\end{equation}
The details of each interaction will be described in Sec.~\ref{sec:components}. A prominent exception to the linear representation Eq.~\eqref{eq:lin-eff-field} is for example the cubic magnetic anisotropy, which can, however, be included approximately using a linearization. Next to cubic anisotropy, other interactions such as the Dzyaloshinskii-Moriya interaction can also be included in the same way. However, in the present work, we limit ourselves to the ones mentioned above, with the main focus being on the dipolar interaction.

Once the magnetization field $\bm{m}$ is not in a minimum of the free energy $\mathcal{G}$, its temporal evolution is given by the Landau-Lifshitz-Gilbert equation of motion
\begin{equation}\label{eq:torque-eq}
    \frac{\mathrm{d}\bm{m}}{\mathrm{d}t} = \omega_M \left( \bm{h}_\mathrm{eff}\times\bm{m} \right) + \bm{T}_\mathrm{G}
\end{equation}
with $\omega_M = \gamma\mu_0  M_\mathrm{s}$ being the characteristic magnetic frequency and $\gamma$ the modulus of the gyromagnetic ratio in the material at hand. The equation of motion must be equipped with the boundary condition $   \bm{n}\cdot\nabla \bm{m} = 0$ which is automatically satisfied by most choices of basis functions in a FEM approach. The first term in  Eq.~\eqref{eq:torque-eq} describes the precessional motion around the so-called \textit{effective} field $\bm{h}_\mathrm{eff}$, comprised of the internal and the external fields. This (unitless) effective field is obtained from the Gibb's free energy as the variational derivative,\footnote{The equality between the first and the second line in Eq.~\eqref{eq:full-eff-field} follows from the choice of $\bm{h}_\mathrm{int}$ in Eq.~\eqref{eq:lin-eff-field} and is not valid in general.}
\begin{align}\label{eq:full-eff-field}
\begin{split}
        -\frac{1}{\mu_0 M_\mathrm{s}^2}\frac{\delta\mathcal{G}}{\delta\bm{m}}& =  \bm{h}_\mathrm{ext} + \frac{\bm{h}_\mathrm{int}}{2} + \frac{1}{2}\left(\pdv{}{\bm{m}}\bm{h}_\mathrm{int}\right)\bm{m}\\& =  \bm{h}_\mathrm{ext} - \hat{\mathbf{N}}\bm{m} \\
        &\equiv \bm{h}_\mathrm{eff}.
\end{split}
\end{align}
The second term $\bm{T}_\mathrm{G}$, in torque equation Eq.~\eqref{eq:torque-eq}, represents viscous (Gilbert) damping and relaxes the magnetization to equilibrium after a certain time \cite{gilbertClassicsMagneticsPhenomenological2004}. When searching for the spin-wave dispersion in a certain magnetic body, we are interested only in the small-amplitude conservative dynamics and can therefore neglect this term. The linear damping rates of individual spin-wave modes can also be recovered from their spatial profiles \cite{verbaDampingLinearSpinwave2018}. To map the spin-wave spectrum, Eq.~\eqref{eq:torque-eq} can be solved using a numerical time integration, as it is done in many micromagnetic codes \cite{donahueOOMMFUserGuide1999,kakaySpeedupFEMMicromagnetic2010, vansteenkisteDesignVerificationMuMax32014}. However, as discussed above, it is desirable to obtain stationary small-amplitude solutions (the spin-wave modes) directly from an eigenvalue problem.

Following \textit{e.g.} Refs.~\citenum{daquinoComputationMagnetizationNormal2012,naletovIdentificationSelectionRules2011, verbaCollectiveSpinwaveExcitations2012,tarasshevchenkonationaluniversityofkyiv64volodymyrskastr.kyiv01601ukraineSpinWavesArrays2013}, the nonlinear torque equation Eq.~\eqref{eq:torque-eq} can be linearized in the usual way, by separating the time-dependent magnetization into a static and a dynamic part according to
\begin{equation}\label{eq:linear-ansatz}
    \bm{m} =  \bm{m}_0(\bm{r})+   \delta \bm{m}(\bm{r},t) + \mathcal{O}(\abs{\delta\bm{m}}^2),
\end{equation}
with $\bm{m}_0(\bm{r})$ being the normalized direction of the equilibrium magnetization and $\delta\bm{m}(\bm{r},t)$ being the dimensionless small-amplitude variation from the equilibrium. Since the norm of the total magnetization vector is constrained by $\abs{\bm{m}}=1$, the equilibrium direction and the dynamical component must be orthogonal, $\bm{m}_0 \hspace{-2pt}\perp\hspace{-2pt}\delta \bm{m}$. Inserting the separation Eq.~\eqref{eq:linear-ansatz} into the torque equation Eq.~\eqref{eq:torque-eq} and neglecting all nonlinear terms, and finally expanding the variable magnetization into linear waves according to
\begin{equation}\label{eq:expansion1}
   \delta\bm{m}(\bm{r},t) =\sum\limits_\nu \bm{m}_{\nu} (\bm{r})\, e^{- i\omega_{\nu} t}
\end{equation}
leads to the linearized Landau-Lifshitz equation
\begin{equation}\label{eq:lin-ll}
 \frac{\omega_{\nu}}{\omega_M} \bm{m}_{\nu} = i\bm{m}_0 \times \vu{\Omega} \bm{m}_{\nu} \quad\text{with}\quad  \bm{m}_0 \hspace{-2pt}\perp\hspace{-2pt}\bm{m}_\nu,
\end{equation}
which is a standard eigenvalue problem that yields the spin-wave mode frequencies $\omega_\nu$ as well as their spatial profiles $\bm{m}_\nu$. It is clear from the demand that $\delta\bm{m}$ is a real vector, that if $\omega_\nu$ is an eigenvalue of Eq.~\eqref{eq:lin-ll} with corresponding eigenvector $\bm{m}_\nu$, then so is $-\omega_\nu$ with eigenvector $\bm{m}_\nu^*$. Here, the asterisk denotes complex conjugation.

The Hamiltonian operator $\hat{\mathbf{\Omega}}$ is a tensor operator given by
\begin{equation}
    \vu{\Omega}(\bm{r}) = h_0 \hat{\mathbf{I}} + \hat{\mathbf{N}},
\end{equation}
here $h_0 = \bm{m}_0 \cdot \bm{h}_\mathrm{eff}(\bm{m}_0)$ is the projection of the unitless static effective field (including any static external field) onto the equilibrium direction $\bm{m}_0$ and  $\hat{\mathbf{I}}$ is the identity operator. The operator $\hat{\mathbf{\Omega}}$ is self-adjoint and, as long as $\bm{m}_0$ is a local minimum of the Gibb's free energy $\mathcal{G}$, it is also positive definite for vectors $\bm{m}_\nu$ perpendicular to $\bm{m}_0$. As a consequence, one can show that all eigenfrequencies $\omega_\nu$ are real valued and, moreover, eigenvectors for different eigenvalues satisfy orthogonality relations \cite{daquinoComputationMagnetizationNormal2012,naletovIdentificationSelectionRules2011}. Let us note, that we write the linearized equation Eq.~\eqref{eq:lin-ll} in a similar form as \textit{e.g.} Refs.~\citenum{naletovIdentificationSelectionRules2011, verbaCollectiveSpinwaveExcitations2012,tarasshevchenkonationaluniversityofkyiv64volodymyrskastr.kyiv01601ukraineSpinWavesArrays2013} with the only difference that our operator $\hat{\mathbf{\Omega}}$ is made dimensionless.

\subsection{Reduction to waveguide cross section}

In order to solve numerically the eigenvalue problem defined in  Eq.~\eqref{eq:lin-ll}, typically, the whole magnetic volume needs to be discretized. Depending on the implementation of the dipolar interaction one even needs to model a large "air box" outside of the specimen. The problem can be significantly simplified when studying the spin-wave spectrum in an extended waveguide with translationally invariant equilibrium along the length of the waveguide. In other words, the equilibrium $\bm{m}_0(\bm{\rho})$ does not change along the propagation direction of the spin waves which we choose to be $\bm{e}_z$ in the lab system, with $\bm{\rho}$ being a spatial vector in the $xy$ plane. Let us stress the point, that the equilibrium magnetization can be inhomogeneous within the waveguide cross section, \textit{i.e.} depend on the coordinates $\bm{\rho}$.
We consider a \textit{long} waveguide with constant cross-section area $A$ and length $L\rightarrow\infty$ (see Fig.~\figref{fig:schematics}{a}). To avoid inflating notation, we denote a set as well its Hausdorff measure (\textit{e.g.} the 2D measure \textit{area} for $A$) with the same symbol. In every instance, the difference will be clear from the context.

\begin{figure}[h!]
    \centering
    \includegraphics[scale=1]{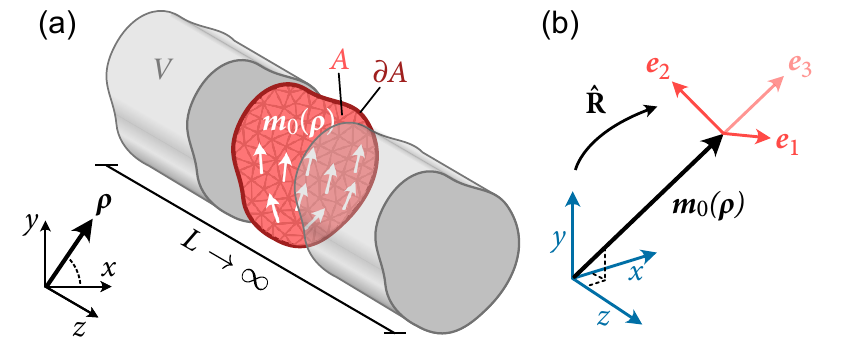}
    \caption{(a) Schematics of an infinitely extended waveguide with arbitrary cross section $A$ and translationally invariant equilibrium magnetization $\bm{m}_0$. (b) Illustration of the transformation $\hat{\mathbf{R}}$ from lab system $\{x,y,z\}$ into local coordinate system $\{\bm{e}_1,\bm{e}_2,\bm{e}_3\}$ attached to the equilibrium magnetization.}
    \label{fig:schematics}
\end{figure}

When letting the length of the waveguide $L$ go to infinity, one can transform the linearized equation into a plane-wave problem and solve it only in a single cross section of the waveguide. Of course, for this case, the Gibb's free energy $\mathcal{G}$ of the whole magnetic element diverges. Thus, the equilibrium configuration $\bm{m}_0(\bm{\rho})$ can be obtained by minimizing the energy density over length $g(\bm{m}_0) = \mathcal{G}(\bm{m}_0)/L$ which remains finite, as $L\rightarrow\infty$.

The spin-wave mode profiles about such a translationally invariant equilibrium can be expressed as %
\begin{equation}\label{eq:lin-ll-plane-wave}
    \bm{m}_\nu (\bm{r}) = \bm{\eta}_{\Tilde{\nu}k}(\bm{\rho})\,e^{ikz}
\end{equation}
with $k$ being the wave number in $z$-direction, $\bm{\eta}_{{\nu} k}$ the lateral mode profiles  which are complex vector fields in the $xy$ plane. Here, $\Tilde{\nu}$ is the lateral mode index. For simplicity, we will drop the tilde again and, from now on, use $\nu$ only as the lateral mode index. Using this definition, Eq.~\eqref{eq:lin-ll} becomes
\begin{equation}\label{eq:llg-plane}
\frac{\omega_{\nu k}}{\omega_M} \bm{\eta}_{\nu k} = i\bm{m}_0 \times \vu{\Omega}_k \bm{\eta}_{\nu k} \quad\text{with}\quad  \bm{m}_0 \hspace{-2pt}\perp\hspace{-2pt}\bm{\eta}_{\nu k},
\end{equation}
with the plane-wave Hamiltonian operator
\begin{equation}
    \hat{\mathbf{\Omega}}_k = e^{-ikz}\,\hat{\mathbf{\Omega}}\, e^{ikz}
\end{equation}
which is the original operator transformed into the waveguide cross section. Clearly, this operator is still self-adjoint. Moreover, if $\hat{\mathbf{\Omega}}$ is positive definite it follows that $\hat{\mathbf{\Omega}}_k$ is too, \textit{i.e.} for some arbitrary lateral profile $\bm{v} = e^{-ikz}\bm{w}$ which satisfies $\bm{v}\perp\bm{m}_0$ (with $\bm{w}$ being the corresponding full volumetric profile)
\begin{equation}
    \langle \bm{v}, \hat{\mathbf{\Omega}}_k \bm{v}\rangle_A = \lim\limits_{L\rightarrow\infty} \frac{1}{L} \langle \bm{w}, \hat{\mathbf{\Omega}} \bm{w}\rangle_V > 0
\end{equation}
with respect to the $\mathbb{L}^2(S)$ scalar product
\begin{equation}
    \langle \bm{a}, \bm{b}\rangle_S= \int_S \mathrm{d}S\, \bm{a}^*\cdot \bm{b}, \qquad S = A, V.
\end{equation}
Therefore, the spectral properties of the initial eigenvalue problems are recovered. By solving Eq.~\eqref{eq:lin-ll-plane-wave} numerically with the wave vector $k$ as a parameter, one obtains the dispersion $\omega_\nu(k) \equiv \omega_{\nu k}$ of different mode branches $\nu$ within the waveguide at hand. Due to the complex eigenmode ansatz proportional to $\exp[{i(kz - \omega_{\nu k}t)}]$, again, all lateral mode profiles come in complex pairs. If $\bm{\eta}_{\nu k}$ is the corresponding eigenvector to the physical eigenvalue $\omega_{\nu k}$, then $\bm{\eta}_{\nu k}^*$ is the eigenvector to the conjugate eigenvalue $-\omega_{\nu,-k}$. In other words, the graphs $(k,\omega_{\nu k})$ of the individual dispersion branches $\nu$ are, in general, point symmetric with respect to $(0,0)$.

\subsection{Rotated eigenvalue problem}

Recall that we want to solve the plane-wave eigenvalue problem Eq.~\eqref{eq:llg-plane} for eigenvectors orthogonal to the equilibrium magnetization, $\bm{m}_0\hspace{-2pt}\perp\hspace{-2pt}\bm{\eta}_{\nu k } $. The implementation of this constraint can be achieved either by introducing an operator which projects vectors into the plane locally transverse to the equilibrium direction $\bm{m}_0$ or by rotating the eigenvalue problem into the local reference frame of $\bm{m}_0$ and reducing the dimension of the eigenvectors \cite{daquinoComputationMagnetizationNormal2012}. The rotation is performed using a suitable unitary transformation $\hat{\mathbf{R}}$ into a local orthonormal system $\{\bm{e}_1,\bm{e}_2,\bm{e}_3\}$ where $\bm{m}_0 = \bm{e}_3$ everywhere (see Fig.~\figref{fig:schematics}{b}). Such a basis is for example given by
\begin{equation}
    \bm{e}_3 = \bm{m}_0, \quad \bm{e}_2 = \frac{\bm{e}_z\times\bm{m}_0}{\vert \bm{e}_z\times\bm{m}_0 \vert }, \quad \bm{e}_1 = -  \frac{\bm{m}_0\times\bm{e}_2}{\vert \bm{m}_0\times\bm{e}_2 \vert }
\end{equation}
By left-multiplying Eq.~\eqref{eq:llg-plane} with this operator $\hat{\mathbf{R}}$ and using its unitarity, \textit{i.e.}  $\hat{\mathbf{R}}^\dagger \hat{\mathbf{R}} = \hat{\mathbf{I}}$, one obtains the rotated eigenvalue problem
\begin{equation}\label{eq:dyn-eigenvalueproblem}
    \frac{\omega_{\nu k }}{\omega_M} \Tilde{\bm{\eta}}_{\nu k } = \hat{\mathbf{D}}\Tilde{\bm{\eta}}_{\nu k }
\end{equation}
with the rotated eigenvectors $\Tilde{\bm{\eta}}_{\nu k } = \hat{\mathbf{R}}{\bm{\eta}}_{\nu k }$, the dynamic-matrix operator
\begin{equation}
    \hat{\mathbf{D}} = i\hat{\bm{\Lambda}}  \hat{\mathbf{R}} \vu{\Omega}_k \hat{\mathbf{R}}^\dagger
\end{equation}
and the dagger denoting Hermitian conjugation.
Due to the rotation of the mode profiles into the local reference frame of $\bm{m}_0$, one realizes that, under the constraint that the mode profiles must be locally perpendicular to the static magnetization, their third component must vanish everywhere, \textit{i.e.} $\Tilde{\bm{\eta}}_\nu \cdot \bm{e}_3 \equiv 0$. This allows to reduce the dimensionality of the profiles from three-dimensional to two-dimensional vectors. This transformation can be performed using the matrix operator
\begin{equation}
   \underline{ \hat{\mathbf{R}}} = \begin{pmatrix}
    \bm{e}_1 \cdot \bm{e}_x & \bm{e}_1 \cdot \bm{e}_y & \bm{e}_1 \cdot \bm{e}_z \\ \bm{e}_2 \cdot \bm{e}_x & \bm{e}_2 \cdot \bm{e}_y & \bm{e}_2 \cdot \bm{e}_z
    \end{pmatrix},
\end{equation}
which is unitary for vectors orthogonal to $\bm{m}_0$. In the local reference frame, the cross product $\bm{m}_0\times ...$ takes the simple form of a multiplication with the matrix (written in $\{\bm{e}_1, \bm{e}_2\}$ basis)

\begin{equation}
    \underline{\hat{\bm{\Lambda}}}= \begin{pmatrix} 0 & -1  \\ 1 & 0\end{pmatrix}.
\end{equation}

\section{Components of the plane-wave Hamiltonian operator}\label{sec:components}

After the plane-wave eigenvalue problem has been formulated, we can introduce the components of the plane-wave operator
\begin{equation}
    \Hat{\mathbf{\Omega}}_k = h_0  \Hat{\mathbf{I}} +   \Hat{\mathbf{N}}_k
\end{equation}
by obtaining the plane-wave versions $\Hat{\mathbf{N}}_k$ of the magnetic tensor for the different interactions considered. For each interaction, the contribution to the equilibrium-field projection $h_0$ can of course be obtained by applying the operator $\Hat{\mathbf{N}}_k$ to the equilibrium configuration $\bm{m}_0$ at $k=0$. For example, if only an external field and exchange interaction would be considered, one would have
\begin{equation}
    h_0 = \bm{m}_0 \cdot (\bm{h}_{\mathrm{ext}} - \hat{\mathbf{N}}^{\mathrm{(exc)}}_0\bm{m}_0).
\end{equation}

\subsection{Uniaxial crystal anisotropy}

The uniaxial magnetocrystalline anisotropy arising from spin-orbit coupling of the spin system with the crystal lattice can be expressed as
\begin{equation}
    \hat{\mathbf{N}}^{(\mathrm{uni})} = - \frac{2K_\mathrm{u}}{\mu_0 M_\mathrm{s}^2}\, \bm{e}_\mathrm{u} \otimes \bm{e}_\mathrm{u},
\end{equation}
with $K_\mathrm{u}$ being the first-order uniaxial-anisotropy constant, $\bm{e}_\mathrm{u}$ being the normalized direction of anisotropy and $\otimes$ denoting the tensor product. Since this interaction does not involve any spatial derivatives the resulting Hamiltonian operator $\hat{\mathbf{\Omega}}^{(\mathrm{uni})}_k$ does not change its form when transformed to the waveguide cross section, \textit{i.e.} $\hat{\mathbf{N}}^{(\mathrm{uni})} = \hat{\mathbf{N}}_k^{(\mathrm{uni})}$ is not dependent on the wave vector.

\subsection{Symmetric exchange interaction}

The exchange interaction within the continuum limit is written as
\begin{equation}
    \hat{\mathbf{N}}^{(\mathrm{ex})} = -\lambda_\mathrm{ex}^2\nabla^2.
\end{equation}
with $\lambda_\mathrm{ex}=\sqrt{2 A_\mathrm{ex}/\mu_0 M_\mathrm{s}^2}$ being the exchange length, $A_\mathrm{ex}$ the exchange stiffness constant of the material and $\nabla^2$ the Laplacian acting on three-dimensional vector fields in three-dimensional space. Projecting the operator into the wave-guide cross section yields
\begin{equation}
    \vu{N}_k^{(\mathrm{ex})} =  e^{-ikz}\vu{N}^{(\mathrm{ex})}e^{ikz}= \lambda_\mathrm{ex}^2 k^2 \vu{I} -  \lambda_\mathrm{ex}^2 \nabla_{\rho}^2.
\end{equation}
Here, $\nabla_{\rho}^2$ denotes the part of the Laplacian operator which operates on a three-dimensional vector fields in the $xy$ plane.

\subsection{Dipolar interaction}

For an arbitrary magnetic specimen, the dipolar interaction can be expressed using the so-called magnetostatic potential $\phi(\bm{r})$ as
\begin{equation}\label{eq:dip-field}
    \hat{\mathbf{N}}^{(\mathrm{dip})} \bm{m} = -\bm{h}_\mathrm{dip} = \nabla \phi.
\end{equation}
The magnetostatic potential is obtained by solving the Poisson equation
\begin{equation}
    \nabla^2\phi(\bm{r},t) = \begin{cases}
    \nabla \bm{m}(\bm{r},t)  & \quad \text{for}\ \bm{r}\in V\\
    0 & \quad \text{elsewhere}.
    \end{cases}
\end{equation}
with appropriate continuity and jump conditions at the boundary $\partial V$ (see for example Eqs.~(4--5) in Ref.~\citenum{fredkinHybridMethodComputing1990}).
For the potential of an individual spin wave with mode profile according to Eq.~\eqref{eq:lin-ll-plane-wave}}, we employ the ansatz
\begin{equation}
    \phi_{\nu k}(\bm{r},t) = \psi_{\nu k}(\bm{\rho})e^{i(kz-\omega_{\nu k} t)}
\end{equation}
with the complex lateral potential $\psi_{\nu k}(\bm{\rho})$. Inserting this ansatz into the initial Poisson equation, we obtain
\begin{equation}\label{eq:screened-poisson}
    \left(\nabla_{\rho}^2 -k^2\right)  \psi_{\nu k}= \begin{cases}
    \left(\nabla_{\rho} +ik\bm{e}_z\right) \bm{\eta}_{\nu k} & \quad \text{for}\ \bm{\rho}\in A\\
    0 & \quad \text{elsewhere}.
    \end{cases}
\end{equation}
Equation \eqref{eq:screened-poisson} is a so-called screened Poisson equation (or Yukawa equation) which, analytically, is equivalent to the inhomogeneous Helmholtz equation by making the substitution $k\rightarrow ik$. We discuss the numerical solution of this equation in more detail in Sec.~\ref{sec:num-implentation}. Once the lateral potential $\psi_{\nu k}$ of a given mode is calculated, the action of the plane-wave dipolar operator on a lateral mode profile is given by
\begin{equation}
    \vu{N}_k^{(\mathrm{dip})} \bm{\eta}_{\nu k} = \left(\nabla_\rho + ik\bm{e}_z\right)\psi_{\nu k}.
\end{equation}

\section{Numerical implementation and the Fredkin-Koehler method for plane waves}\label{sec:num-implentation}

For a numerical solution of the eigenvalue problem in Eq.~\eqref{eq:dyn-eigenvalueproblem}, the equilibrium magnetization $\bm{m}_0(\bm{\rho})$, mode profiles $\bm{\eta}_{\nu k}(\bm{\rho})$,  the potentials $\psi_{\nu k}(\bm{\rho})$ as well as all involved operators in the dynamic matrix $\Hat{\mathbf{D}}$ can be discretized within the cross section $A$ of the waveguide by projecting them onto triangular elements with linear shape functions. Here, we use a mass-lumping technique to obtain the discretized differential operators. After discretization, all operators including $\nabla$ or the transformation $\hat{\mathbf{R}}$ (except the dipolar operator) take the form of sparse matrices and their actions are simply matrix-vector multiplications. Therefore, the numerical implementation can be kept close to the mathematical formulation. Once the dynamic matrix is constructed, it can be diagonalized using standard iterative eigensolvers such as the Arnoldi-L\'anczos method\cite{lanczosIterationMethodSolution1950,arnoldi1951principle}.

The evaluation of the dipolar interaction requires special care. The lateral potential $\psi_{\nu k}$ in the dipolar operator $\Hat{\mathbf{N}}^\mathrm{(dip)}$ could be readily obtained by the convolution of the right-hand side of the screened Poisson equation Eq.~\eqref{eq:screened-poisson} with the appropriate Green's function
 \begin{equation}
     G_k(\bm{\rho},\bm{\rho}^\prime) = -\frac{1}{2\pi}K_0(\abs{k}\abs{\bm{\rho} -\bm{\rho}^\prime})
 \end{equation}
using the modified Bessel function of second kind and zeroth order $K_0$. On a discrete mesh with $n$ nodes, this approach requires the computation of $\mathcal{O}(n^2)$ matrix elements of the Green's function \cite{fredkinHybridMethodComputing1990}. Moreover, due to the logarithmic singularity at $\abs{\bm{\rho} -\bm{\rho}^\prime}=0$ and the rapid decay for $\abs{\bm{\rho} -\bm{\rho}^\prime}>0$ of the Green's function, it is hard to evaluate the convolution integral without adding up a considerable amount of noise.

An alternative of lower complexity and much higher precision is to solve for the potential directly using a hybrid boundary-element/finite-element method, commonly referred to as the Fredkin-Koehler method~\cite{fredkinHybridMethodComputing1990}, which only requires computing $\mathcal{O}(n^{4/3})$ matrix elements. In the following, we will expand this method to plane waves. Let us note that, in the finite-difference case of rectangular elements, the plane-wave dipolar tensor can be obtained directly as a dense matrix, as was done for the first time by Henry \textit{et al.} in Ref.~\citenum{henryPropagatingSpinwaveNormal2016}.

Typically, if one is to solve for the potential using FEM, the boundary conditions for the potential at infinity need to be specified. To avoid having to model a large air box, these boundary conditions can be mapped onto the cross-section boundary $\partial A$ using the Fredkin-Koehler method for plane waves. This allows to solve for the potential by integrating over the sample cross-section area, only.

Within this section, we drop the indices $\nu$ and $k$ for the sake of visual clarity. The potential subject to the screened Poisson equation Eq.~\eqref{eq:screened-poisson} has to be equipped with the following continuity and jump conditions at the boundary of the magnetic element
\begin{equation}\label{eq:continuity}
    \eval{\psi_\mathrm{out}(\bm{\rho})}_{\partial A} - \eval{\psi_\mathrm{in}(\bm{\rho})}_{\partial A} = 0
\end{equation}
and
\begin{equation}\label{eq:normal-derivative}
    \eval{\pdv{\psi_\mathrm{out}(\bm{\rho})}{\bm{n}(\bm{\rho})}}_{\partial A} - \eval{\pdv{\psi_\mathrm{in}(\bm{\rho})}{\bm{n}(\bm{\rho})}}_{\partial A} = - \eval{\bm{n}(\bm{\rho})\cdot\bm{\eta}(\bm{\rho})}_{\partial A}
\end{equation}
where $\psi_{\mathrm{in}}$ and $\psi_{\mathrm{out}}$ are the lateral potential inside and outside of the magnetic specimen, respectively. These two conditions follow directly from the properties of the full volumetric potential $\phi(\bm{r},t)$ when the surface-normal vector field $\bm{n}(\bm{\rho})\perp\bm{e}_z$. Furthermore, we require that the potential vanishes as $\abs{\bm{\rho}}\rightarrow\infty$.

\begin{figure}
    \centering
    \includegraphics{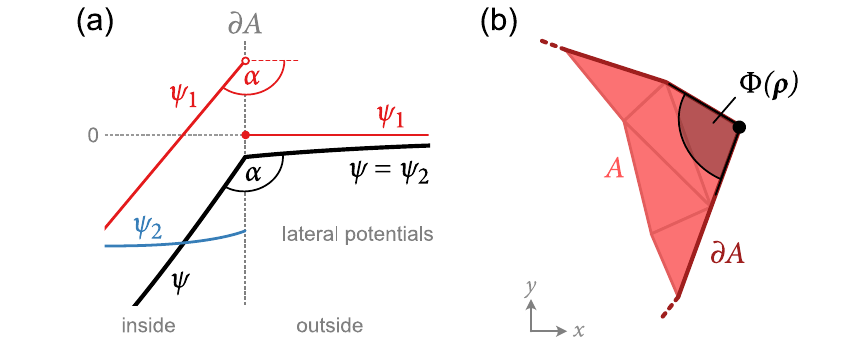}
    \caption{(a) Schematics of the different lateral potentials in the Fredkin-Koehler method for plane waves, shown inside and outside of the magnetic sample when crossing the boundary $\partial A$. For simplicity, we only show the real parts of the potentials here. (b) Definition of the angle subtended at a certain boundary node point.}
    \label{fig:fredkinkoehler}
\end{figure}
The motivation behind the Fredkin-Koehler method is to map the Dirichlet boundary conditions at infinity to the boundary of the magnetic sample by splitting the potential into two parts $\psi = \psi_1 + \psi_2$. The first potential $\psi_1$ satisfies an inhomogeneous Neumann problem inside the magnetic volume and is zero outside of the specimen such that it produces the jump in the normal derivative according to Eq.~\eqref{eq:normal-derivative}. The second potential $\psi_2$ solves a homogeneous Dirichlet problem with boundary conditions obtained from $\psi_1$ at the boundary $\partial A$ such that it compensates the jump of $\psi_1$ and therefore guarantees the continuity of the full lateral potential $\psi$ according to Eq.~\eqref{eq:continuity} (see Fig.~\figref{fig:fredkinkoehler}{a}). In summary, we have the system of equations
\begin{align}
  &  \begin{rcases}
 (\Delta - k^2)\psi_1 & = (\nabla + ik\bm{e}_z)\bm{\eta} \ \ \\ (\Delta - k^2)\psi_2 & = 0
\end{rcases}
&\text{\ in $A$}&\\
&\qquad\  \begin{rcases}
 \pdv{}{\bm{n}}\psi_1 & = \bm{n}\cdot\bm{\eta}\qquad\quad\ \ \ \\ \psi_2 &= u(\bm{\rho})
\end{rcases}
&\text{\ at $\partial A$}&\\
&\begin{rcases}
\qquad\quad\ \ \ \, \psi_1 & = 0 \qquad\qquad\quad\ \\
(\Delta - k^2)\psi_2 & = 0
\end{rcases}
&\text{\ outside $A$}&
\end{align}
Here, $u(\bm{\rho})$ is the Dirichlet boundary condition for $\psi_2$. After having calculated $\psi_1$ numerically using FEM, this boundary condition can be obtained by
\begin{equation}\label{eq:dirichlet-bc}
\begin{split}
        u(\bm{\rho}) & = \frac{1}{2\pi} \int\limits_{\partial A} \mathrm{d}s^\prime\, \psi_1(\bm{\rho}^\prime) \ \pdv{}{\bm{n}} K_0(\abs{k}\abs{\bm{\rho} -\bm{\rho}^\prime}) \\ & \qquad + \left( \frac{\Phi(\bm{\rho})}{2\pi} - 1\right) \psi_1(\bm{\rho}), \quad\bm{\rho}\in \partial A
\end{split}
\end{equation}
Here $\Phi(\bm{\rho})$ is the angle subtended by the boundary point $\bm{\rho}$ within an arbitrary cross section (see Fig.~\figref{fig:fredkinkoehler}{b}), which is $\pi$ for a smooth boundary and $\dd{s^\prime}$ is the line element on $\partial A$. The Dirichlet boundary condition in Eq.~\eqref{eq:dirichlet-bc} is the plane-wave version of the initial relation for the regular three-dimensional Poisson problem (equation 14 in the seminal paper by Frekdin and Koehler \cite{fredkinHybridMethodComputing1990}) and is rigorously derived in the supplementary material). In a numerical implementation, it can be expressed using a dense-matrix multiplication in the sense
\begin{equation}
    \underline{\psi_2}= \underline{\Hat{\mathbf{B}}_k}\; \underline{\psi_1}
\end{equation}
where $\underline{\psi_{1,2}}$ are the mesh vectors of the two potentials at the boundary and $\underline{\Hat{\mathbf{B}}_k}$ is a dense matrix. Above, we retained the index $k$ at the Green's function $G_k(\bm{\rho},\bm{\rho}')$ and the dense matrix $\underline{\Hat{\mathbf{B}}_k}$ to highlight the fact that both are wave-vector dependent. Let us note, that this method only requires to calculate $\mathcal{O}(n_B^2)$ matrix elements per wave vector (with $n_B$ being the number of boundary nodes) instead of $\mathcal{O}(n^2)$ when storing the whole Green's function. As a result, the presented method is most effective for cross-section meshes with a large surface-to-boundary-node ratio.

A note has to be made on the numerical evaluation of the boundary integral in Eq.~\eqref{eq:dirichlet-bc}. As a specific property of the inhomogeneous Neumann problem for $\psi_1$, the first potential in the plane-wave Fredkin-Koehler method diverges as $k\rightarrow 0$ (see supplementary material). This divergence is counteracted by the second potential $\psi_2$ such that the total potential $\psi$ remains finite. Thus, the correct treatment of the dipolar interaction in our method is very sensitive on the proper calculation of the boundary values, \textit{i.e.} on the dense matrix $\underline{\Hat{\mathbf{B}}_k}$. For this reason, we recommend at least a sixth-order segmentation of the boundary elements to evaluate the boundary integral, weighted linearly on the $2^6 +1$ resulting nodes. Due to the reasonably small number of boundary nodes and elements the dense matrices are in general small in storage requirements and can be computed on the fly for the different $k$ values.

\section{Applications}\label{sec:applications}

To showcase and validate our method, we calculate the spin-wave dispersion for three different examples and compare the results for different methods or theoretical predictions typically used to investigate spin-wave dynamics. We do not intend to discuss the physical implications of the different examples in detail, but rather highlight the capabilities of our numerical approach.

\subsection{Longitudinally magnetized rectangular waveguide}

\begin{figure}[h!]
    \centering
    \includegraphics{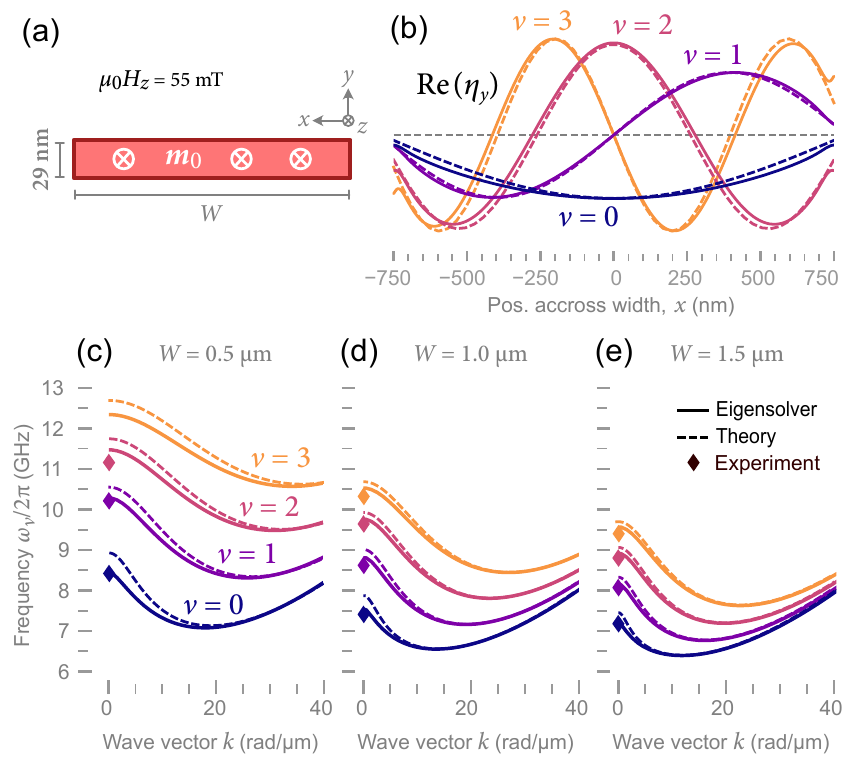}
    \caption{(a) Schematics of the longitudinally magnetized rectangular waveguide overlayed with the equilibrium magnetization. (b) Numerically and theoretically calculated spatial mode profiles of the first four dispersion branches $\nu$ at $k=0$ for a width of the waveguide $W=\SI{1.5}{\micro\meter}$, shown as line scans across the width of the waveguide. (c-e) Dispersion of the same branches for different widths of the waveguide, obtained using our eigensolver (solid lines) and compared with theoretical prediction (dashed) calculated according to Refs.\citenum{kalinikosTheoryDipoleexchangeSpin1986,guslienkoEffectiveDipolarBoundary2002a} and experimental data (rhombus) from Ref.~\citenum{roussigneExperimentalTheoreticalStudy2001}.}
    \label{fig:bv-stipe}
\end{figure}

As a first example, we calculate the spin-wave dispersion for a standard magnonic problem, namely a rectangular waveguide with fixed thickness $T=\SI{29}{\nano\meter}$ and different widths $W$\!, where the equilibrium magnetization $\bm{m}_0$ is aligned along the propagation ($z$) direction (see Fig.~\figref{fig:bv-stipe}{a}). In such a case, the spin waves exhibit a negative slope of the dispersion $\omega_\nu(k)$ for small wave vectors, and are typically referred to as backward-volume modes. In the case of a waveguide with finite width, the dispersion is split into multiple branches $\nu$, which belong to standing waves along the width of the waveguide (see Fig.~\figref{fig:bv-stipe}{b}). For small aspect ratios $T/W\ll 1$, their dispersion is analytically described using the theory of Kalinokos and Slavin \cite{kalinikosTheoryDipoleexchangeSpin1986} with effective dipolar pinning conditions introduced by Guslienko \textit{et al.} \cite{guslienkoEffectiveDipolarBoundary2002a}. These pinning conditions lead to an \textit{effective} width of the waveguide $W_\mathrm{eff}$ which is typically larger than the physical width $W$, resulting in increased wavelengths along the width $\lambda_\nu = 2W_\mathrm{eff}/(\nu+1)$. To compare our results with the experiments of Roussign\'e \textit{et al.} \cite{roussigneExperimentalTheoreticalStudy2001}, we apply an external field of $\mu_0 H_z = \SI{55}{\milli\tesla}$, set a saturation magnetization of $M_\mathrm{s}=\SI{621}{\kilo\ampere\per\meter}$ and a reduced gyromagnetic ratio of $\gamma/2\pi = \SI{29.76}{\giga\hertz\per\tesla}$. The exchange stiffness is set to a typical value for Ni$_{80}$Fe$_{20}$ (permalloy) of $A_\mathrm{ex}=\SI{13}{\pico\joule\per\meter}$. The triangular mesh used has an average edge length of \SI{5}{\nano\meter}.

In Figs.~\figref{fig:bv-stipe}{c-e}, we show the dispersion calculated using our eigensolver for different widths of the waveguide $W$ for the lowest four branches $\nu=0,1,2,3$, overlayed with theoretical and experimental data. Overall, we achieve a very good agreement with the experimental data from Ref.~\citenum{roussigneExperimentalTheoreticalStudy2001}. As expected, the agreement with the theoretical calculations (performed according to Refs.~\citenum{kalinikosTheoryDipoleexchangeSpin1986,guslienkoEffectiveDipolarBoundary2002a}) increases as the aspect ratio of the waveguide decreases. For the smallest aspect ratio, \textit{i.e.} the largest width $W=\SI{1.5}{\micro\meter}$, we show the real part of the numerically calculated out-of-plane component $\Re(\eta_z)$ of the spatial mode profiles at $k=0$ as line scans along the width ($x$) direction, together with the theoretical prediction.

\subsection{Edge shape in transversally magnetized rectangular waveguide}

\begin{figure}[h!]
    \centering
    \includegraphics{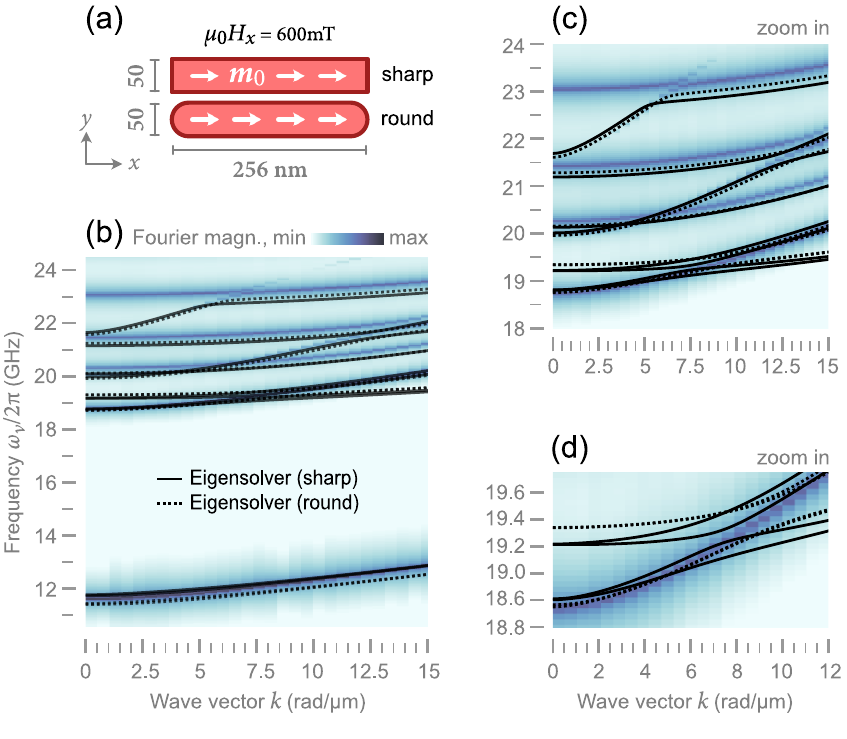}
    \caption{(a) Schematics of the two different waveguide cross sections overlayed with the equilibrium magnetizations. (b) Dispersion map (wave-vector- and frequency-dependent Fourier magnitude) for the waveguide with sharp edges obtained with MuMax$^3$ and shown as a colormap. Overlayed are the dispersion curves for sharp (solid) and round (dotted) corners, obtained with our eigensolver. (c,d) show zoom-ins of the same data, highlighting the avoided level crossings of the different dispersion branches as well as the influence of the edge roundness on the nature of the crossings.}
    \label{fig:de-stripe}
\end{figure}

After having showcased a regular rectangular cross section, we now present the transition to curved geometries. For this purpose we calculate the spin-wave dispersion for a transversally magnetized waveguide of \SI{50}{\nano\meter} thickness and \SI{256}{\nano\meter} width, once with sharp corners and once with round corners, as shown in Fig.~\figref{fig:de-stripe}{a}. In the case of a transversally magnetized stripe, the equilibrium magnetization $\bm{m}_0$ is perpendicular to the propagation direction and the spin waves, also referred to as magnetostatic surface waves, exhibit a positive dispersion at small wave numbers.

We set the saturation to $M_\mathrm{s}=\SI{796}{\kilo\ampere\per\meter}$, exchange stiffness to $A_\mathrm{ex}=\SI{13}{\pico\joule\per\meter}$ and reduced gyromagnetic ratio to $\gamma/2\pi = \SI{28}{\giga\hertz\per\tesla}$. An external field of $\mu_0 H_x = \SI{600}{\milli\tesla}$ is applied along the width ($x$) direction. At this field, the waveguide is almost completely saturated except for small regions at the sides (known as flower state). The equilibrium distribution $\bm{m}_0(\bm{\rho})$ is found by minimizing the energy length density $g(\bm{m}_0)=\mathcal{G}(\bm{m}_0)/L$ using a conjugate-gradient method.

We compare our results for the waveguide with sharp edges to a numerically calculated dispersion assuming the same material parameters, obtained using the GPU accelerated micromagnetic solver MuMax$^3$ \cite{vansteenkisteDesignVerificationMuMax32014} which employs a finite-difference method to solve the Landau-Lifshitz-Gilbert equation Eq.~\eqref{eq:torque-eq} on a rectangular grid. Spin waves were excited using an out-of-plane oscillating field which is homogeneous along the width (for details, see supplementary material). As a consequence, only modes with an even (symmetric) spatial profile along the width can be excited.
Modelling a waveguide with round edges is computationally much more expensive in a finite-difference code such as MuMax$^3$ since it requires a much finer discretization along the thickness of the sample. In our finite-element dynamic-matrix approach, both cross sections require almost the same computational effort.

In Fig.~\figref{fig:de-stripe}{b}, we show the Fourier magnitude $P(k,\omega)$ obtained with MuMax$^3$ as a heatmap, overlayed with the the corresponding dispersions (sharp and round edges) obtained with our eigensolver. Considering the different techniques of discretization, we achieve a very good agreement between both methods. Notably, multiply avoided level crossings between the various dispersion branches are recovered (see also Figs.~\figref{fig:de-stripe}{c}). We see that round edges at the waveguide sides mainly result in a shift of the overall frequencies which can be a blue- or a red shift depending on the branch. However, as seen in \figref{fig:de-stripe}{d}, round edges can have a dramatic impact on the avoided crossings between the branches.

Let us repeat here an another benefit of the propagating-wave dynamic-matrix approach over the full three-dimensional approach used by MuMax$^3$.~Since, in the latter case, wave vectors are obtained by means of spatial Fourier transform, the length of the waveguide has to be increased in order to achieve more wave-vector precision. Depending on the problem at hand, this may result in considerably more computational time. In the plane-wave approach used here, the wave vector simply appears only as a parameter in the resulting eigenvalue problem and, thus, can be varied continuously.

\subsection{Magnetic nanotubes with easy-plane anisotropy}

\begin{figure}[h!]
    \centering
    \includegraphics{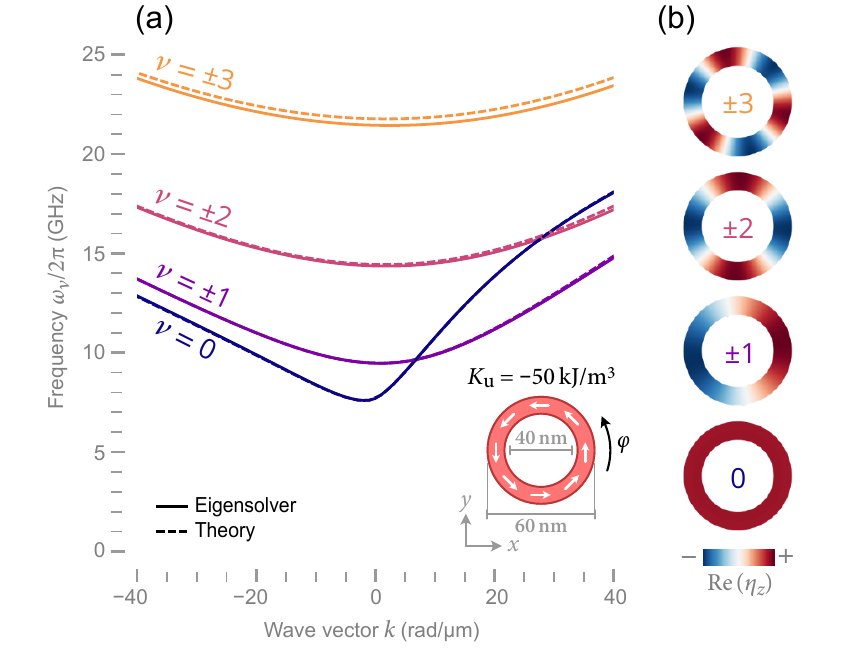}
    \caption{(a) Numerically (solid) and theoretically (dashed) obtained asymmetric spin-wave dispersion in a vortex-state magnetic nanotube with easy-plane anisotropy. The branches correspond to the first five azimuthal mode indices around $\nu=0$, while modes with opposite index $\pm\nu$ are degenerate. The inset shows the cross section of the nanotube overlayed with the equilibrium magnetization. (b) Numerically obtained spatial mode profiles (here only the real part of the $z$ component $\eta_z$ is shown).}
    \label{fig:tube}
\end{figure}

As a last example, we go fully into the field of curvilinear magnonics and calculate the spin-wave dispersion of a magnetic nanotube with thin mantle magnetized in the vortex state (see the inset in Fig.~\figref{fig:tube}{a}). We consider a tube with \SI{60}{\nano\meter} outer and \SI{40}{\nano\meter} inner diameter. At such a size and in the absence of any external field, a vortex state can be stabilized using an easy-plane anisotropy which is, in our case, a uniaxial anisotropy along the $z$ axis with negative constant $K_\mathrm{u}=\SI{-50}{\kilo\joule/\cubic\meter}$. Other than this, we use the same \textit{permalloy} material parameters as in the previous example.

The spin-wave dispersion in such a system exhibits a curvature-induced asymmetry ($\omega_\nu(k) \neq \omega_\nu(-k)$) of purely dipolar origin (see Ot\'alora \textit{et al.} \cite{otaloraCurvatureInducedAsymmetricSpinWave2016, otaloraAsymmetricSpinwaveDispersion2017}). For the case of a thin mantle, the lateral profiles can be described approximately as homogeneous along the thickness $\rho$ and being proportional to $\bm{\eta}_\nu \propto \exp(i\nu\varphi)$ with $\varphi$ being the azimuthal angle with respect to the $z$ axis. Here, the lateral mode index $\nu$ counts the periods along the azimuthal direction and can take positive and negative values. In the vortex state, modes with opposite $\nu$ are degenerate, \textit{i.e} their frequencies are equal $\omega_\nu = \omega_{-\nu}$. In Fig.~\figref{fig:tube}{a}, we compare the asymmetric dispersion calculated using our eigensolver with the theoretical predictions by Ot\'alora, showing an almost perfect agreement. Deviations in the higher-order azimuthal modes $\nu=\pm 3$ are of exchange origin and can be decreased by further decreasing the discretization of the mesh, which in our case was \SI{3}{\nano\meter}. Finally, Fig.~\figref{fig:tube}{b} shows the real part of $z$ component $\Re(\eta_z)$ of the numerically obtained spatial mode profiles which correspond, as expected, to harmonic waves running in azimuthal direction. Whereas, here for the case of tubes with thin mantle, our numerical approach serves to confirm a theoretical model, it may become useful \textit{e.g.} for the study of magnetic nanotubes with thick mantle for which theoretical models are not available at the time and full three-dimensional dynamic simulations become computationally demanding.

As a comment on the performance of our propagating-wave finite-element approach: We obtained the same dispersion as in Fig.~\figref{fig:tube}{a} -- without being able to separate degenerate branches -- using the highly-optimized GPU-accelerated three-dimensional finite-element code TetraMag \cite{kakaySpeedupFEMMicromagnetic2010}. This was done by using a mesh with similar number of elements within the cross section, but extruded to a certain length in propagation direction to give an adequate wave-vector resolution. Even with the efficient field-pulse method described in the supplementary material and exciting several azimuthal branches up to $\abs{\nu}=3$ at the same time the computation (including post-processing) took several days on a high-performance TITAN Xp GPU. Using the also GPU-accelerated finite-difference code MuMax$^3$ requires a similar amount of time due to the large number of rectangular cells needed in the cross section to adequately model the curved surface of the tube. For comparison, obtaining the dispersion and the corresponding mode profiles with a higher wave-vector resolution using an unoptimized implementation of our finite-element propagating-wave dynamic-matrix approach on a standard laptop CPU took less than 15 minutes.

\section{Conclusion}\label{sec:conclusion}

In summary, we have presented a finite-element propagating-wave dynamic-matrix approach to efficiently calculate the spin-wave dispersion in waveguides with arbitrary (bounded) cross section and translationally invariant magnetic equilibrium along the propagation direction. This was achieved by numerically solving a plane-wave version of the linearized equation of motion of the magnetization. In contrast to dynamic micromagnetic simulations, spin-wave frequencies and mode profiles are obtained without any post-processing. Important characteristics of the spin waves such as linear damping rate or dynamic susceptibility can be calculated from the spatial mode profiles (see \textit{e.g.} Ref.~\citenum{verbaDampingLinearSpinwave2018} and Ref.~\citenum{verbaCollectiveSpinwaveExcitations2012}, respectively). Our approach differs from the finite-difference approach presented in Ref.~\citenum{henryPropagatingSpinwaveNormal2016} mainly in the discretization type and in the way the dipolar potential of the spin-wave modes is calculated at a given wave vector. For this purpose we presented an extension to the Fredkin-Koehler method to the screened Poisson equation for propagating waves, which is suitable to model waveguides with arbitrarily shaped bounded cross section. For a number of known systems, our results were validated using theoretical predictions, dynamic micromagnetic simulations and experimental results. Although, in the magnetic interactions considered, we have restricted ourselves to uniaxial anisotropy as well as dipolar and exchange interaction, other contributions like the Dzyaloshinskii-Moriya interaction can be included in the same way. Due to the possibility to model waveguides with surface curvature, we believe that our approach is of particular relevance for the emerging field of curvilinear magnetism but also useful for standard problems in magnetization dynamics.

\section*{Supplementary material}

See supplementary material which contains the derivation of the Dirichlet boundary condition Eq.~\eqref{eq:dirichlet-bc} within the Frekdin-Koehler method for plane waves. Moreover, we show how the divergence of the first lateral potential $\psi_1$ is canceled out by the second one $\psi_2$. Finally, the dynamic micromagnetic simulations used for comparison in Sec.~\ref{sec:applications} are described.

\section*{Acknowledgements}

The authors are very thankful to Burkhard Clau\ss, Steffen Boerm and Jorge A. Ot\'alora for fruitful discussions. Financial support by the Deutsche Forschungsgemeinschaft within the programs KA 5069/1-1 and KA 5069/3-1 is gratefully acknowledged.

%

\end{document}